\input{aipcheck}

\documentclass[
    ,final            
  ]
  {aipproc}

\layoutstyle{6x9}

\begin{document}

\title{Broad-band electromagnetic radiation from microquasars interacting 
      with ISM}

\author{V. Bosch-Ramon}{
  address={Departament d'Astronomia i Meteorologia, 
  Universitat de Barcelona, Av. Diagonal 647, 
  E-08028 Barcelona, Catalonia (Spain)},
  email={vbosch@am.ub.es},
}

\author{F. A. Aharonian}{
  address={Max-Planck-Institut fur Kernphysik, 
  Saupfercheckweg 1, Heidelberg, 69117, Germany},
}
 
\iftrue
\author{J. M. Paredes}{
  address={Departament d'Astronomia i Meteorologia, 
  Universitat de Barcelona, Av. Diagonal 647, 
  E-08028 Barcelona, Catalonia (Spain)},
}

\begin{abstract}
Microquasars (MQs) are galactic objects with relativistic jets that constitute a
source population which can be responsible for production of a non-negligible
fraction of the observed galactic cosmic rays. These relativistic protons, 
associated with the termination of the jet, interact with the interstellar
medium and, at certain surrounding conditions, may lead to production of
detectable fluxes of high-energy and very high-energy gamma-rays. 
This radiation is accompanied by
the broad-band emission of secondary electrons from decays of 
$\pi^\pm$-mesons produced through synchrotron, bremstrahlung and inverse
Compton process. The features of broad-band emission initiated by 
proton-proton (pp)
interactions in such a scenario is discussed in the context of the strategy of
search for counterparts of high-energy and very high-energy gamma-ray sources in the
galactic plane.    
\end{abstract}

\maketitle


\section{Introduction}

In this paper we explore a scenario where microquasar jets  
(see e.g. \cite{mirabel}) initiate indirect and persistent sources of gamma-rays
through interactions of high energy protons, accelerated by jet termination
shocks, with high density regions of the interstellar medium (ISM). We study the
multiwavelength properties of such sources using a model that calculates the
broadband spectrum of the emission coming out from the decay products of {\em
pp}  interactions -  $\pi^0$-mesons which lead to direct gamma-ray emission and
charged $\pi^\pm$-mesons which initiate broad-band emission through
synchrotron, inverse Compton and bremsstrahlung losses of  secondary electrons.
This model takes into account the propagation effects due to the energy
dependent diffusion of protons  during their travel from the accelerator to the
cloud  (for more details, see \cite{bosch1}). Below we discuss two
cases relevant to both impulsive MQs, i.e. X-ray binaries with transient jets,
and continuous MQs, i.e. X-ray binaries  with persistent jets.

\section{MQ-cloud interactions and production of gamma-rays}

The jets of MQs end somewhere within the Galaxy, although it is still unclear
the way they terminate (\cite{heinz1}, \cite{heinz2}). Assuming that the
jet contains a significant population of protons and that that these protons
are (re)accelerated at the terminal part of the jet to very high energies,
interactions between these  particles and nearby dense gas clouds leads to the
production of neutral and charged pions.  Then, $\pi^0$-mesons  decay to
gamma-ray photons and $\pi^{-/+}$-mesons decay to e$^-$ and e$^+$. The
secondary electrons radiatively cool due to interactions with the ambient 
magnetic and radiation fields and gas. Typically, synchrotron emission of these
electrons extends from radio frequencies to X-rays and bremstrahlung leads to
high energy radiation  from X-rays to TeV gamma-rays (provided that the protons
are accelerated at least to energies 100 TeV). One may expect also gamma-rays
of inverse Compton origin, but in most cases the contribution of this component
is quite small.  

Due to energy-dependent propagation effects, the broad-band spectral energy
distribution ($\nu F_\nu$ or $\epsilon L_\epsilon$) of radiation initiated by
protons strongly depends  on the age, the temporal character (impulsive or
continuous) of the accelerator and the distance between the accelerator (site
of the jet termination)  and the gas target, e.g. a dense molecular cloud
(\cite{aharonian}). The source is expected to be extended with a
characteristic size of the dense cloud, persistent and steady (on time scales
$\gg 1$ yr). It should be noted  that hadronic gamma-rays could be produced at
much smaller scales as well, e.g. at interaction of the relativistic jet with
the dense wind of the stellar companion (\cite{romero1}).

\section{Results}

Below we present the results of numerical calculations performed  for the model
parameters summarized in Table 1. The most important  parameters  are the
diffusion coefficient of cosmic rays, assumed in the form  $D(E)=10^{27} (E/1 \
\rm GeV)^{0.5} \ \rm cm^2/s$,  the magnetic field  ($B=5 \times 10^{-4} \ \rm
G$) in the gamma-ray production region,  as well as the  acceleration spectrum
of protons which is  assumed to be a power-law with spectral index $p=2$ with
an exponential cutoff $E_{\rm p, max}=10^5  \  \rm GeV$.  Absolute fluxes of
radiation  are  determined by the  density of the ambient gas ($n=10^4 \ \rm
cm^{-3}$)  and the total energy released in protons in the case of impulsive 
accelerator, $E_{\rm k}$, or kinetic energy luminosity in protons in the case 
of continuous accelerator, $L_{\rm k}$.    

  
\begin{table}
\begin{tabular}{lrr}
\hline
\tablehead{1}{l}{b}{Parameter}
  & \tablehead{1}{r}{b}{Symbol}
  & \tablehead{1}{r}{b}{Value}   \\
\hline
Diffusion coefficient normalization constant 
& $D_{10}$ & $10^{27}$~cm$^2$~s$^{-1}$\\
Diffusion power-law index 
& $\chi$ & 0.5\\
ISM medium density 
& $n$ & 0.1~cm$^{-3}$\\
High density ISM region/cloud density 
& $n_{\rm H}$ & 10$^4$~cm$^{-3}$\\
Mass of the high density ISM region/cloud 
& $M$ & $5\times10^3~M_{\odot}$\\
Magnetic field within the cloud 
& $B$ & $5\times10^{-4}$~G\\
IR radiation energy density within the cloud 
& $U$ & 10~eV~cm$^{-3}$\\
Planckian grey body temperature (IR) 
& $T$ & 50~K\\
Power-law index of the high energy protons 
& $p$ & 2\\
Cut-off energy of the high energy protons 
& $E_{\rm p max}$ & 10$^5$~GeV\\
Kinetic energy luminosity for protons in the continuos MQ 
& $L_{\rm k}$ & 10$^{37}$~erg/s\\
Kinetic energy  for protons in the impulsive MQ 
& $E_{\rm k}$ & 10$^{48}$~erg\\
\hline
\end{tabular}
\caption{The basic model parameters.}
\label{constants}
\end{table}

In Figs. 1 and 2, we show the broad-band spectral energy distribution (SED) of
radiation of a molecular cloud at  different epochs of observations ($t=0$
corresponds to the start of operation of the accelerator) for the case of
impulsive MQ and continuous MQ assuming that the cloud is located at a distance
$R=10$ pc from the  accelerator. One can see significant differences between
the SEDs  corresponding to a sudden release of protons (Fig.~1) and  a constant
injection of protons (Fig.~2). In particular, in the  case of continuous MQ we
see a quite narrow distribution of $\pi^0$-decay and synchrotron radiation;
this reflects the fact that at any given time and at the fixed distance from
the accelerator the protons have a narrow distribution with characteristic
energy which decreases with time. In the case of continuous MQ the spectra of
protons inside the cloud have broader distributions, especially at later
epochs, which transfers to the secondary electromagnetic radiation. 

\begin{figure}
\caption{The spectral energy distribution (SED) of radiation of a 
dense cloud under bombardment of  protons accelerated at the 
termination  of an  impulsive MQ  at a distance $R$=10~pc. 
The results are shown for three different epochs: 
for 100~yr (curve 1),  for 1000~yr (2),  and for 10000~yr (3). 
The model parameters are described in Table 1.}
\includegraphics[height=.25\textheight]{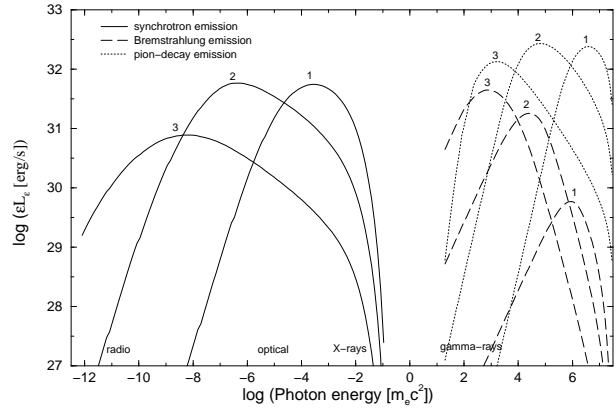}
\end{figure}

\begin{figure}
\caption{The same as in Fig.~1 but for a continuous MQ for the 
time epochs  $t$=100~yr (1), 1000~yr (2), and 10000~yr (3)).}
\includegraphics[height=.25\textheight]{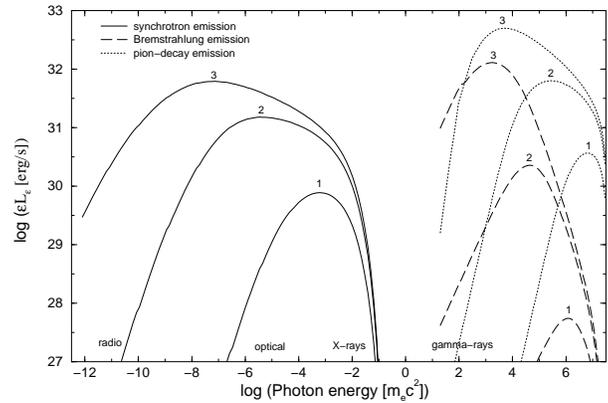}
\end{figure}

In Figs. 3 and 4, we show the broad-band SEDs at the fixed epoch, $t=1000 \ \rm
yr$, but at different distances to impulsive and the continuous MQs, 
respectively (note that the luminosity axis range is not like in Figs. 1 and
2). These figures demonstrate the effect of propagation, namely  how the higher
energy particles reach earlier the target, with its impact  on the secondary
electromagnetic emission. 

\begin{figure}
\caption{The SEDs of radiation of clouds located 
at three different distances $R=$~100~pc (curve 1),  
30~pc (2), and 10~pc (3) from an impulsive MQ at $t$=1000~yr.}
\includegraphics[height=.25\textheight]{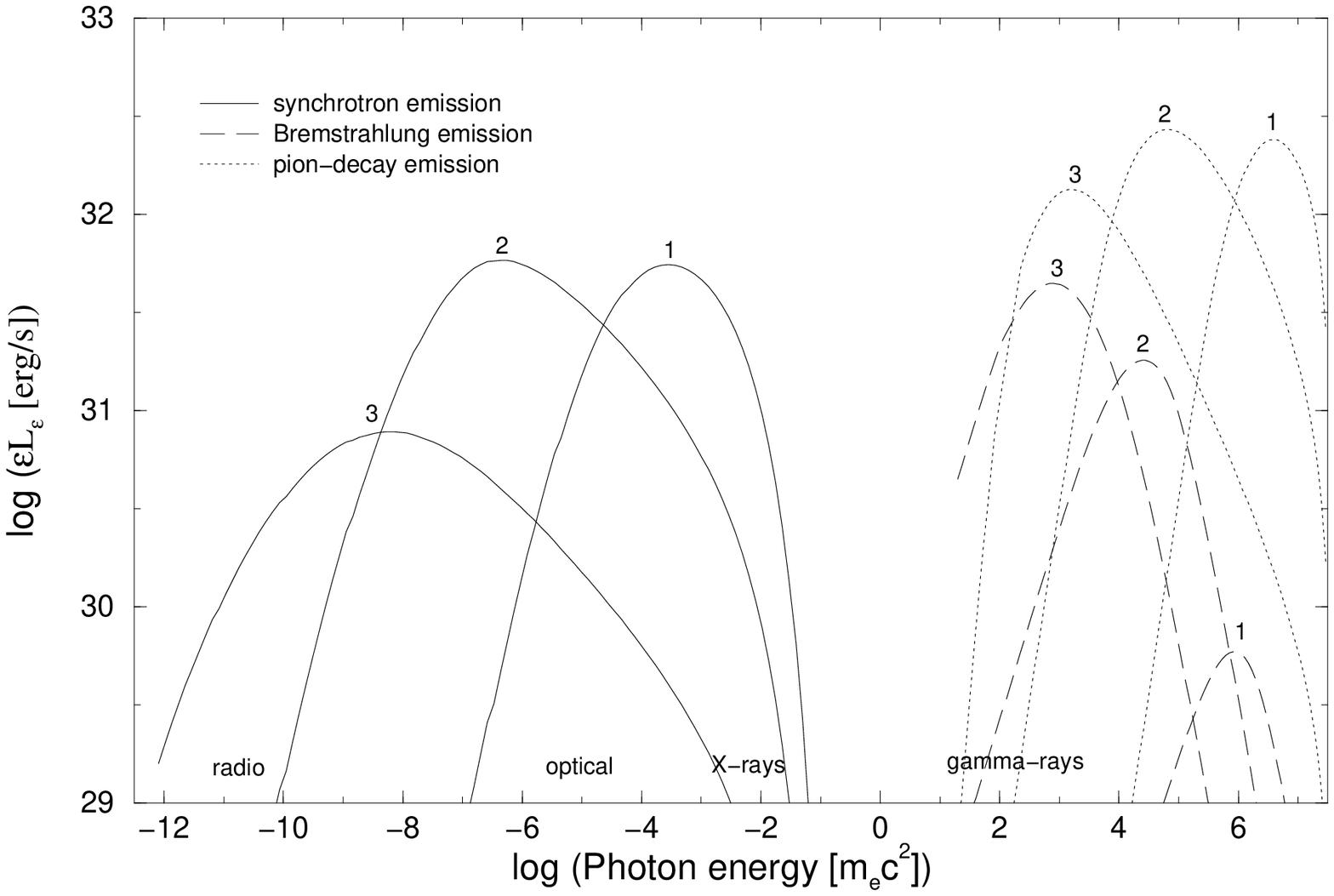}
\end{figure}

\begin{figure}
\caption{The same as in Fig.~3
but for a continuous MQ:  $R$=100~pc (curve 1), 
30~pc (2), and 10~pc (3)).}
\includegraphics[height=.25\textheight]{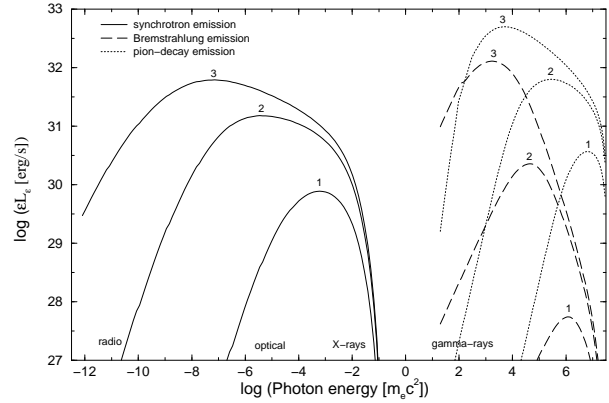}
\end{figure}

\begin{figure}
\caption{
SED of radiation of a molecular cloud  located at a distance of 10 pc from a
position of termination of the continous MQ jet ($t$=10$^5$~yr) with  a
recent short-term activity which took place 200 yr ago. The radiation
components related to the continuous jet are shown  by dashed lines;  the
components related to the burst type event are shown by dotted lines. 
The superposition of all radiation components is shown
by the solid curve.
}
\includegraphics[height=.25\textheight]{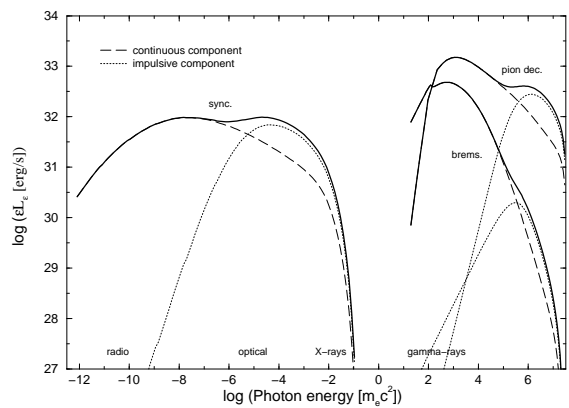}
\end{figure}

MQs show variability on different timescales, which implies that 
proton luminosity should be treated as a mean value.
Moreover, these objects often undergo higher states of activity, and therefore
a strong outburst with release of large amount of  relativistic  protons could
have happened recently.  If so, one  should expect  a mixture of ``old'' CRs
accumulated  over large timescales, typicall $10^5 - 10^6$ years,  and
``young'' CRs  associated with  a recent active phase of the MQ jet. In  this
case we should expect  characteristic spectral features of both  persistent  
and burst type accelerators as demonstrated  in Figs. 1 and 2. In Fig. 3 we
show an example of such a spectrum  initiated by a quasi-continuous accelerator
of age $t=10^5$ year with a mean proton luminosity $10^{37} \ \rm erg/s$ 
(dashed lines) on top of the contribution  from a recent, 200 year old, 
burst-like  event with total energy release in relativistic protons  $10^{48} \
\rm erg$. The superposition of these two contributions (solid lines) results in
a quite unusual  spectral shape over a very large energy region from radio to
TeV energies. While the TeV gamma-rays and  X-rays are associated  with
interactions of potons produced during the recent burst type event,  MeV/GeV
gamma-rays and synchrotron radiation at infrared and longer wavelengths  are 
contributed mainly by the ``old'' proton population  produced before the last
burst. Detection of such spectra from specific dense regions of the
interstellar medium could be  an indicator of presence of a nearby persistent
accelerator, e.g. a MQ jet, which recently was in an active phase.   

While the search of  such objects in TeV gamma-rays will constitute an 
important part of the future surveys of the galactic plane by forthcoming
ground-based  instruments,  it is possible that some such objects are already
detected by EGRET. In fact, statistical studies of the unidentified EGRET
sources in the galactic plane point to a possible link with high density
regions in the inner spiral arms (\cite{bhattacharya}). It is likely that
the low latitude EGRET sources are  grouped in two subpopulations; one formed
by  variable sources  with possible association with MQs (\cite{bosch2}), 
and another subpopulation consisting of  persistent objects (\cite{romero2}).  
The results of this study show that a part of  the persistent 
EGRET sources also could be (indirectly) related to MQs through termination of
their jets in dense environments of the interstellar medium. Feasibility of
such a scenario is demonstrated in Fig. 4 where the calculations are compared
with the spectrum and luminosity of a ``standard'' non-variable EGRET source.

\begin{figure}
\caption{The gamma-ray luminosity of a molecular 
cloud of mass 10$^5$~M$_{\odot}$ located at  $R$=10~pc from 
the termination site of a continuous MQ jet. 
The spectrum of a ``standard'' 
unidentified  EGRET source  is also shown (long-dashed line).}
\includegraphics[height=.24\textheight]{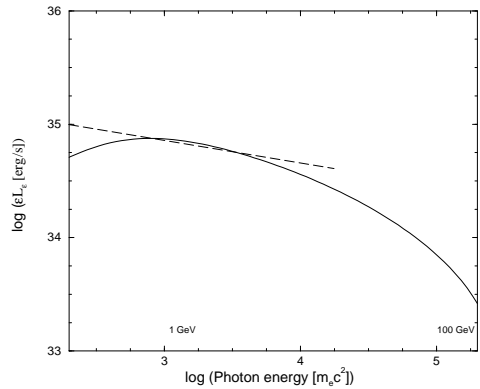}
\end{figure}

\section{Conclusions}

In this work, we have explored the scenario of gamma-ray production due to 
termination of MQ jets in dense environments of the interstellar medium, e.g. 
close to massive molecular clouds. The spectrum of  broad-band radiation
induced by interaction of protons with molecular clouds  depends strongly on
several parameters, in particular on the age of the accelerator, the  distance
accelerator--target, and energy dependent diffusion of protons. In the case of
termination of MQ jets with $L_{\rm k} \sim 10^{37} \ \rm erg/s$ close to dense
clouds  ($n \sim 10^{4} \ \rm cm^{-3}$) extended secondary synchrotron
radiation can be detected from radio to X-ray frequencies. Moreover, gamma-ray 
emission from GeV to TeV energies could be detected by the new generation of
satelite-borne and ground-based gamma-ray instruments like GLAST and HESS. 

\begin{theacknowledgments}
V.B-R. and J.M.P. acknowledge partial support by DGI of the Ministerio de
Ciencia y Tecnología (Spain) under grant AYA-2001-3092, as well as additional
support from the European Regional Development Fund (ERDF/FEDER). During this
work, V.B-R has been supported by the DGI of the Ministerio de Ciencia y
Tecnología (Spain) under the fellowship FP-2001-2699.
\end{theacknowledgments}




{}



\end{document}

\endinput